\documentclass{article}
\usepackage{latexsym,amssymb}
\usepackage{amssymb,amsmath}
\usepackage[final]{graphicx}
\input epsf
\def\ac{\hfill\break\noindent}
\def\bfa{{\bf a}}\def\bfr{{\bf  r}}
\def\bfm{{\bf  m}}\def\bfk{{\bf k}}\def\bfq{{\bf q}}
\def\ie{{\em i.e.}}\def\eg{{\em e.g.}}
\def\cA{{\cal A}}\def\cZ{{\cal Z}}
\def\vsm{{v_{0}}}\def\V0{V_{\bf 0}}\def\C0{{C_0}}
\def\hH{{\hat H}}\def\hT{{\hat T}}\def\hV{{\hat V}}\def\hnu{{\hat \nu}}
\def\vxi{\vec{\xi}}
\def\tphi2{{\widetilde{{\phi_0}^2}}}\def\tt0{{\widetilde{{\Theta_0}}}}
\def\td{{\tilde {d^2}}}\def\tG2{{\widetilde {G^2}}}

\begin{document}

%
%%%   The headers.
%
%%%   These three macros are to have correct headings in your paper.
%%%   You shall omit all the arguments in the two macros `\euro{}{}{}{}'
%%%   `\Date{}' and fill in `\shorttitle{}'.
%%%   If there is more than one author in the
%%%   \shorttitle macro, use the macro \etal after first author's name
%%%   to obtain the correct heading.
%
%
\title{Strongly localized quantum crystalline states and behaviour
of the dilute jellium model}

\author{Salvino Ciccariello \\
%\institute{
   Dipartimento di Fisica "G. Galilei",\\ Universit\'a di Padova, \\
via Marzolo 8, I-35131 Padova, Italy}
%\pacs{
%\Pacs {61}{10.Eq}{X-ray scattering (including small-angle scattering)}
%\Pacs {61}{12.Ex}{Neutron scattering techniques (including small-angle
%scattering)}
%\Pacs {61}{25.Hq}{Macromolecular and polymer solutions}
%\Pacs {61}{66.Dk}{Alloys}
%  }
\maketitle
\begin{abstract}
We name strongly localized quantum crystalline state (SLQCS) a determinantal
wave-function of the single-particle wave-functions obtained 
by crystalline translations of a wave-function different from zero 
only within a primitive cell of the considered quantum crystalline 
phase.  SLQCSs accurately reproduce the
low density behaviour of the quantum crystals that, as Wigner's
crystalline phase of the jellium model, can become very dilute. Our
analysis explicitly deals with this system. We show that the SLQCS
energy per particle at large dilution ($r_s$) behaves as
$-M_{dl,\sigma}/r_s+C_{\sigma}/{r_s}^{3/2}$, where $M_{dl,\sigma}$
turns out to be the Madelung constant of the considered cubic symmetry
$\sigma$ and $C_{\sigma}$ is a positive constant only numerically
determined. Moreover, as the density gets smaller and smaller, each
electron becomes more and more confined to the centre of its cell.
\vskip 2.truecm
PCAS: 05.30.-w, 71.10.Ca, 71.15.Nc, 73.20.Qt\\
\\

%\rightline{DFPD/08/Th/06\quad\quad\quad\quad\quad}
\hfill{DFPD/08/Th/06}

\end{abstract}
\vfill\eject
%{} {} {} {} {} {}  {} {} {} {} {} {}  {} {} {} {} {} {}

\section{Introduction}
The {\em jellium} model \cite{Fetter, Mahan} is the simplest model of metallic
conductors. It approximates the crystalline array of positive ions  by a
classical, uniform, positive charge density while the
valence electrons are described as quantum mechanical particles interacting
among themselves and  with  the electric field of the aforesaid neutralizing
charge distribution. In this way, the
Hamiltonian of the system, confined at first within a box of volume $V$, is
\begin{eqnarray}
\hH &=& \frac{e^2n^2}{2}\int_V dv\int_V dv'
\frac{e^ {-\mu\vert\bfr - \bfr'\vert}} {\vert\bfr - \bfr'\vert}
-e^2 n\sum_{i=1}^N\int_V dv \frac {e^ {-\mu\vert\bfr_i - \bfr\vert}}
{\vert\bfr_i - \bfr\vert}\label{1.1}\\
\quad & &-\sum_{i=1}^N \frac {\nabla_i^2} {2m}
 + \frac{e^2}{2} \sum_{1\le i\ne j\le N}
\frac {e^ {-\mu\vert\bfr_i - \bfr_j\vert}} {\vert\bfr_i - \bfr_j\vert}.
\nonumber
\end{eqnarray}
Here, $N$ denotes the number of the electrons contained in the box,
$n\equiv N/V$ the electron  number density, $\bfr_i$ the position vector
of the $i$th electron, $e$ and $m$ electrons' charge and mass. The cut-off
parameter $\mu$ goes to zero after taking the thermodynamic limit:
$V\to\infty$, $N\to\infty$ with $n$ fixed. Besides, in Eq.~(\ref{1.1}) and
throughout the paper we adopt units such that $\hbar =1$.

The first order perturbative approximation of the energy of the fundamental
state of $\hH$ was evaluated by Wigner\cite{WigPR} for the unpolarized case
and reads
\begin{equation}\label{1.2}
E_ {u}=   \langle F_u\vert \hH\vert F_u\rangle = N \frac {e^2}{2a_0}
\biggl(\frac{ 2.209}{r_s^2}   -  \frac{ 0.916}{ r_s}\biggr)
\end {equation}
where $\vert  F_u\rangle$ denotes the fundamental state of $N$ free
and unpolarized spin $1/2$ particles, $a_0=1/me^2$ is the Bohr atomic
radius, $r_s\equiv r_0/a_0$ the perturbation parameter with
$r_0\equiv (3/4\pi n)^{1/3}$. The corresponding approximation for the
polarized case was obtained by Bloch\cite{Bloch} and reads
\begin{equation}\label{1.3}
E_{ p} = \langle F_p\vert \hH\vert F_p\rangle  =N \frac {e^2} {2a_0}
\biggl(\frac {3.508}{r_s^2}   -  \frac {1.154} { r_s}\biggr),
\end {equation}
where $\vert  F_p\rangle$ denotes the fundamental state of  $N$ polarized
free fermions. Expressions (\ref{1.2}) and (\ref{1.3}) refer to homogeneous
systems.
Their comparison shows that the polarized configuration is the one stable
at lower density, \ie\ for $r_s > 5.7$. The energy of the classical
crystals, formed by  electrons arranged on a lattice with simple (sc) or
body (bcc) or face centred cubic (fcc) symmetry and a neutralizing
positive uniform background, is $-N(e^2/2a_0)M_{dl,\sigma}/r_s$,
where $M_{dl,\sigma}$ with $\sigma=1,2,3$ denotes the Madelung constant
respectively equal to $1.76012$, $1.79186$ or $1.79175$\cite{Marder} for
the sc, bcc or fcc case. Since at high dilution these energies become
definitely
smaller than Eq.~(\ref{1.3})'s, Wigner  concluded that, at large $r_s$,
the fundamental state of the jellium no longer can be that of the homogeneous
Fermi gas at $T=0^0K$.  It must have a crystalline structure.
In fact, assuming that each electron oscillates around its equilibrium
position and that these positions form a bcc lattice, Wigner\cite{WigTrans}
obtained the following energy expression
\begin{equation}\label{1.4}
E_{crW } =  N \frac {e^2} {2a_0}  \biggl( -\frac{ 1.79186}{r_s} +
\frac{3}{r_s^{3/2} }\biggr),
\end{equation}
where the term proportional to $r_s^{-1}$, representing the energy of the
classical crystal, was put in, so to say, by hand.
A better approximation was later obtained by Carr\cite{Carr} expanding, as
Wigner did, the electrons' position vectors $\bfr_j$ around the bcc lattice
values and, contrarily to Wigner, also retaining the lowest order terms that
couple the resulting harmonic oscillators among themselves.  The resulting
first order approximation of the fundamental state energy is
\begin {equation} \label{1.5}
E_{crC}=  N \frac {e^2} {2a_0}\biggl (-\frac{ 1.79186}{ r_s}+
\frac{2.66}{r_s^{3/2}}\biggr).
\end{equation}
At large $r_s$ we have $E _{crC } < E _{crW } <  E _{p }$. On the basis of
these inequalities one cannot however invoke the Ritz-Rayleigh
principle to conclude that at high dilution  the jellium model is in a bcc
crystalline state because expressions (\ref{1.4}) and (\ref{1.5}) were not
obtained considering the full Hamiltonian $\hH$ of the jellium model, as
defined in Eq.~({1.1}), but two different modifications of it.
The property can however be proved showing that the solution, with the
bcc symmetry, of the jellium's Hartree-Fock (HF) equation yields an energy
lower than Eq.~(\ref{1.3})'s at large $r_s$.  To this aim we recall that
the solution $\psi(\bfr_1,\ldots,\bfr_N)$ of the HF equations is the wave
function that has a determinantal structure and minimizes the expectation
value of $\hH$\cite{Messiah,FoulMiNeRa}. Further, in writing down the
determinantal wave-function, one usually assumes that the single particle
wave-functions have a particular functional form. In particular, assuming
that the single particle wave-functions are the plane waves $e^{i\bfk_j\cdot
\bfr }$ with $j=1,\ldots,N$, the resulting HF energy coincides with
expressions (\ref{1.2}) or (\ref{1.3}) depending on the
considered polarization degree. On the contrary, a HF crystalline
solution can be obtained assuming that the single particle wave-functions
$\varphi_j(\bfr)$ have the form $e^ {i\bfk_j\cdot\bfr }u(\bfr)$, with
$j=1,\ldots,N$ and $u(\bfr)$ a periodic function (with the chosen crystalline
symmetry) to be determined through the minimization of the HF energy.
Applying this procedure, it was recently found\cite{Trail} that
crystalline HF energies become smaller than Eq.~(\ref{1.3})'s at $r_s\ge 4.4$
and that the bcc HF energy is the smallest  in the region $r_s>13.3$. In
this way, Wigner's statement that diluted  jellium assumes a bcc
crystalline configuration is put on a firmer basis because it is now
a consequence of the Ritz-Rayleigh principle. Strictly speaking the stability
of the crystalline phase cannot be considered rigorously proved because the
exact fundamental state is not required to have a determinantal structure,
as assumed in the HF equations. However, the statement's validity
is also supported by Quantum Monte Carlo (QMC) calculations
\cite{CeperAl,OrtiBa,KwoCepMa,OrtHaBa,Zong,DruRaTrToNe} that,
relaxing the determinantal restriction, confirm that
the stable phase of the dilute electron jellium is the bcc one
(even though the exact $r_s$ value of the transition is not yet well
assessed because, depending on the paper, it  ranges from 50 to 105).

The aim of this paper is to present a simpler procedure to
show the existence, for the jellium model, of crystalline phases
more stable than the fluid one and to illustrate how these quantum-mechanical
crystalline solutions reproduce the classical crystal behaviours as
$r_s\to\infty$. These solutions are essentially obtained by a simplified
HF procedure as follows. First, one assumes that the confining box
$V$ consists of $N=(2M+1)^3$ primitive cells with the
chosen crystalline symmetry. Then, one assumes that the single particle
wave-function  $\varphi_j(\bfr,\alpha)$, with $j=1,\ldots,N$, differs from
zero  only within the $j$th primitive cell and goes to zero at the cell's
border. Besides, each $\varphi_j(\bfr,\alpha)$ is obtained
translating a normalized function $\phi(\bfr,\alpha)$
that differs from zero only within $V_{\bf 0}$ (the primitive cell centred
at the origin) and goes to zero at the border
of $V_{\bf 0}$. The corresponding translation vector is
$\sum_{i=1}^3 m_{j,i}\bfa_{i}$,  the $\bfa_{i}$s being the vectors
specifying the primitive cell and ${\bf m}_j=(m_{j,1},m_{j,2},m_{j,3})$
the vector (with integer components) that labels the $j$th cell. Finally,
$\alpha$ is a parameter to be determined by a variational procedure that will
be described later. It must be noted that, if $j\ne k$, functions $\varphi_j
(\bfr,\alpha)$ and  $\varphi_k(\bfr,\alpha)$  are orthogonal because their 
supports have void intersection. The determinantal wave-function is
\begin{equation}\label{1.6}
\psi(\bfr_1,\ldots,\bfr_N,\alpha)=\frac{1}{\sqrt{N!}}
\sum_P (-1)^P \varphi_1(\bfr_{i_1},\alpha)
\varphi_2(\bfr_{i_2},\alpha)\ldots \varphi_N(\bfr_{i_N},\alpha),
\end{equation}
where the sum is performed over all the permutations  $i_1,i_2,\ldots,i_N$
of $1,2,\ldots,N$ and $(-1)^P$ is the parity of the considered permutation.
Hereafter, wave-function (\ref{1.6}) is referred to as a strongly localized
quantum crystalline state (SLQCS) if the involved $\varphi_j(\bfr,\alpha)$s
obey the properties mentioned above. Then, one evaluates the expectation value
of $\hH$ over the above normalized wave-function. In the thermodynamic and
subsequent $\mu\to 0$ limits,  the resulting energy expression
depends on $\alpha$ and $r_s$, and one looks for the $\alpha$ value that
makes the energy value minimum for each $r_s$ value.
In this way one finds (see Fig.1) that:  i) the resulting sc, bcc and fcc
SLQCS energies become smaller than Eq.~(\ref{1.3})'s at $r_s\ge 28$,
$r_s\ge 36$ and  $r_s\ge 38$, respectively, ii) the SLQCS fcc energy nearly
coincides with the bcc one. More definitely, it is always smaller 
than the bcc's  and their slight difference monotonically decreases 
throughout the explored $r_s$ range $[0,10^4]$.  
The sc SLQCS energy at first is smaller that the
fcc's and bcc's and becomes greater of the last two at $r_s\ge500$,
iii) the three SLQCS energies  approach the
relevant Madelung contributions as $r_s\to\infty$. Consequently, the bcc SLQC
energy approaches the corresponding leading term  of Eq.s~(\ref{1.4}) and
(\ref{1.5}),  iv) the optimized $\alpha$ value increases
with $r_s$ and at large $r_s$ one has $\alpha\propto \sqrt{r_s}$,
v)in the same $r_s$ region, the subleading asymptotic terms of the sc, bcc
and fcc SLQCS energies are proportional to $r_s^{-3/2}$, and vi) our SLQCS
energies are always greater than those obtained in Ref.\cite{Trail} that
lie close to the full circles shown in Fig.1a.
{\vskip 0.truecm
{\begin{figure}
{{\leftskip -3.9truecm
\includegraphics[width=0.75\textwidth]{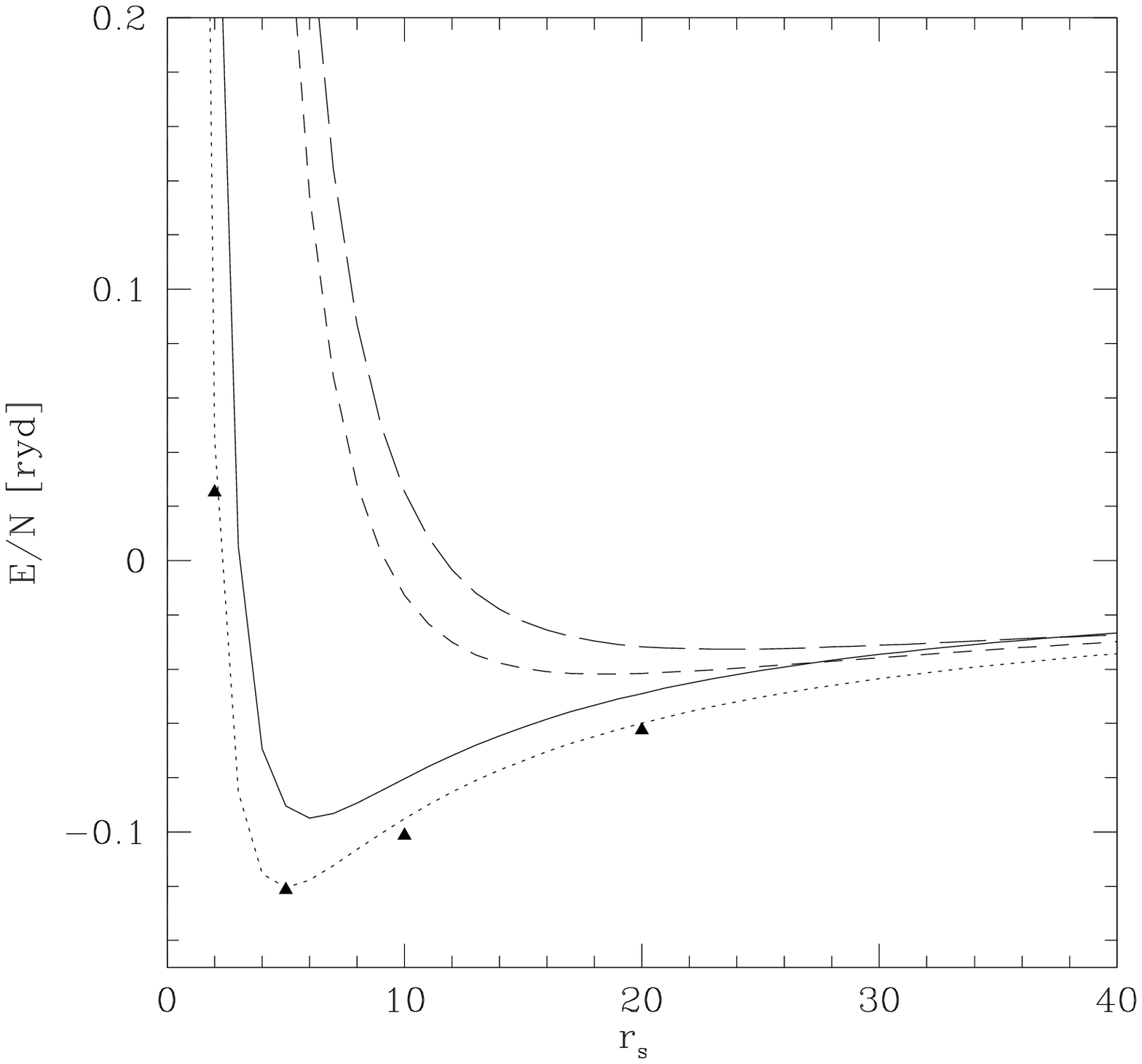}\par
}
\vskip -11.0truecm
{\leftskip 6.truecm
\includegraphics[width=0.75\textwidth]{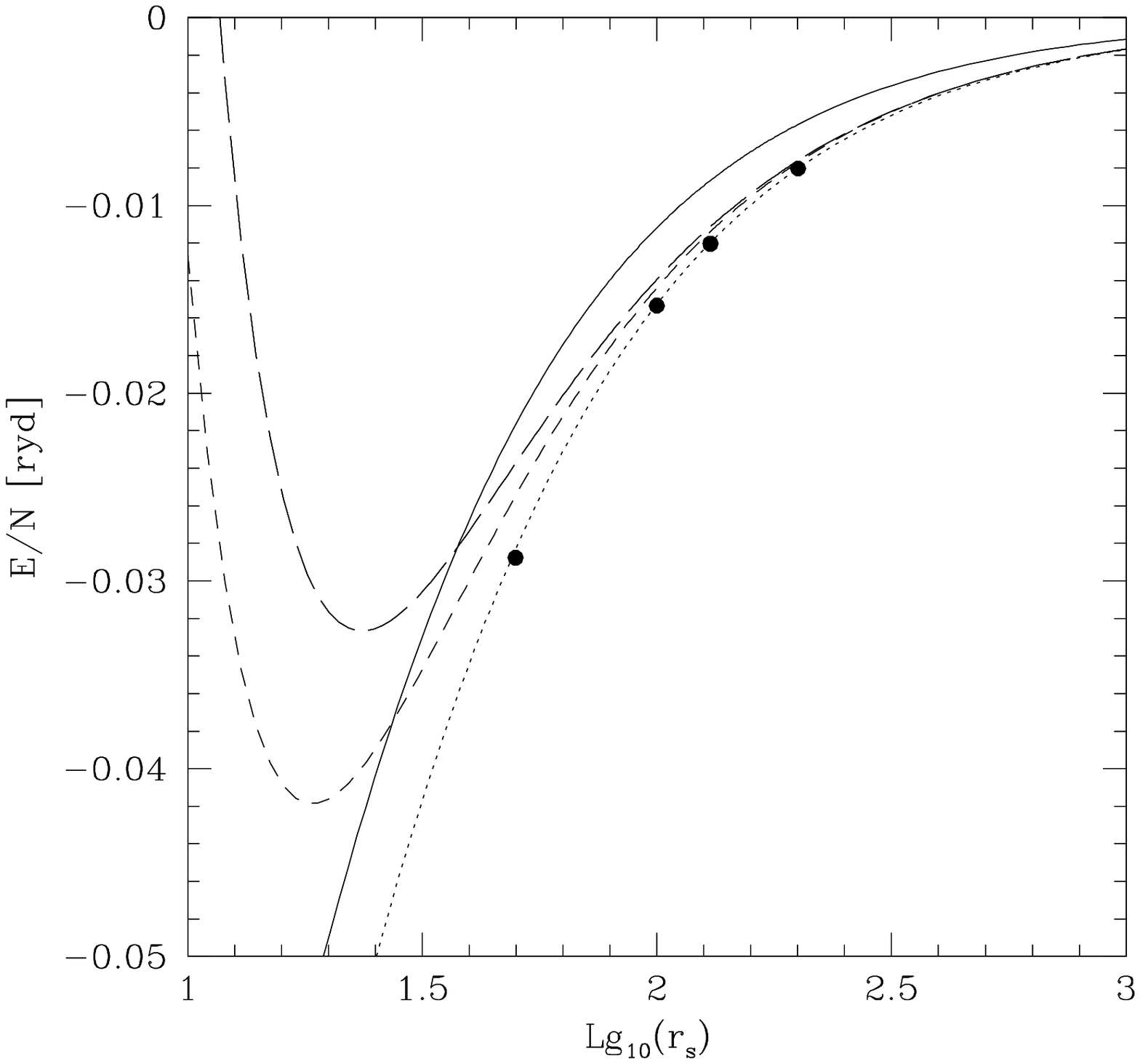}\par
}}
\vskip -3.truecm
{\caption {\label{fig1} Left and right parts: respectively the energy per
particle of the fully polarized jellium model (in Rydberg units) in the inner
and outer part of the $r_s$-range. (Note the $r_s$ logarithmic-scale on the
right.)
In both figures, the continuous lines refer to Eq.~(\ref{1.3}), the dotted
lines to Carr's bcc expression (\ref{1.5}), and the short-dash and 
the long-dash curves to the SLQC solutions with the sc or the bcc symmetry, 
respectively. The full triangles and circles correspond to the QMC
results of Ref.s~\cite{Zong,DruRaTrToNe}. The SLQCS fcc solution is not
shown because it lies very close to the bcc. }}
\end{figure}}
Result vi) is not unexpected because the above HF procedure minimizes
$\langle \psi\vert \hH\vert\psi\rangle$ over a class of functions that,
for the  assumption that the $\varphi_{\jmath}(\bfr,\alpha)$s have compact
support, is more  restricted than that used in Ref.\cite{Trail}.
The SLQCS procedure is however interesting for its numerical simplicity and
for the choice of the single particle wave-functions. The latter's noticed
orthogonality property has an important consequence: if
we evaluate the expectation value of $\hH$ over the normalized Hartree
wave-function
\begin{equation}\label{1.7}
\psi_{H}(\bfr_1,\ldots,\bfr_N,\alpha)=
\varphi_1(\bfr_{1},\alpha)
\varphi_2(\bfr_{2},\alpha)\ldots \varphi_N(\bfr_{N},\alpha),
\end{equation}
we obtain the same expectation value resulting from the fully
antisymmetric state (\ref{1.6}). In other words, SLQCSs have the
property that the Hartree and the Hartree-Fock equations
coincide because the exchange
contributions turn out to be equal to zero as it will be
explicitly shown
below Eq.~(\ref{2.13}). The equivalence of Eq.~(\ref{1.6})
with (\ref{1.7})
in their final results  and the condition that the
$\varphi_j(\bfr,\alpha)$s
vanish on their cell borders imply that in the the SLQCS
approach each electron
in practice is bound to lie within a single cell of the lattice.
This feature on the one hand explains why wave-function (\ref{1.6}) was  
named strongly localizedquantum crystalline state. On the other hand,  
the same feature is shown  by classical ionic crystal. Hence, it is not
surprising that the quantum-mechanical description,
based on  SLCQSs, reproduces the classical crystal behaviour
in the limit $r_s\to\infty$. This appears already evident
from  results iii) and iv). In particular result iv)
implies that, as $r_s \to\infty$, each electron becomes
more and more confined to the centre of the primitive cell
that the electron occupies,   while result iii)
quantum mechanically reproduces the $O(1/r_s)$ term, manually introduced
into Eq.s~(1.4) and (1.5) by Wigner and Carr. \ac
The derivation of these results will now be detailed  according to the
following
plan.  In section 2 we work out the analytical
expressions of the expectation value  $\langle \psi\vert\hH\vert\psi\rangle$
in direct and reciprocal space for the three mentioned cubic symmetries.
In section 3 we specify the functional form of $\phi(\bfr,\alpha)$ and report
the relevant numerical results already anticipated at i)-vi). Section 4
concludes
the paper. Some mathematical details are left to appendices A and B.

\section{Analytical expressions}
The primitive cell $\V0$ centred at the origin is defined as
\begin{equation}\label{2.1}
\V0\equiv\{\bfr\Bigl\vert\ \bfr=\sum_{i=1}^3\frac{\xi_i\bfa_i}{2},\
\-1<\xi_i<1,\ i=1,2,3\},
\end{equation}
where the involved $\bfa_i$s are respectively specified in Eq.s (27a),
(29) and (34) of Chapt. 2 of Ref.~\cite{Kittel} for the sc, bcc
and fcc lattices. For these symmetries the primitive cell volumes are
\begin{equation}\label{2.2}
\vsm=a^3,\ \ a^3/2,\ \ a^3/4,
\end{equation}
$a$ denoting the cubic lattice spacing. The corresponding $r_s$ values are
\begin{equation}\label{2.2a}
r_s=(3/4\pi)^{1/3}a/a_0,\quad (3/8\pi)^{1/3}a/a_0\quad {\rm and}\quad
(3/16\pi)^{1/3}a/a_0.
\end{equation}
We define now the wave-function relevant to a SLQCS. To this aim, we  first
introduce a real even function $\phi(\bfr,\alpha)$ that differs from
zero only within $\V0$ and depends on a positive real parameter $\alpha$.
Whatever $\alpha$, $\phi(\bfr,\alpha)$ and its first and second partial
derivatives are continuous
throughout $\V0$ and vanish as $\bfr$ approaches the border of $\V0$.
Besides, $\phi(\bfr,\alpha)$ is normalized, \ie\
\begin{equation}\label{2.3}
\int_{\V0}\phi^2(\bfr,\alpha)\,dv=1,
\end{equation}
and the $\alpha$ dependence is such that
\begin{equation}\label{2.4}
\lim_{\alpha\to\infty}\phi^2(\bfr,\alpha)=\delta(\bfr),
\end{equation}
$\delta(\bfr)$ denoting the three-dimensional (3D) Dirac function.
Assume now that the box of volume $V$, where the jellium is confined,
contains $(2M+1)$ primitive cells along each direction $\bfa_i$. Then
the jellium will contain $N=(2M+1)^3$ electrons,
$\V0$ is one of the possible cells and $\vsm=V/N$. We denote the
primitive cells either as $V_{\bfm}$ (with $-M\le m_{i}\le M$ and
$i=1,2,3$) or, after performing a convenient relabeling
$\bfm\rightarrow\jmath(\bfm)$, as $V_{\jmath}$ with
$\jmath=1,\ldots,N.$ The inverse relabeling exists and reads
$j\to {\bfm}(j)$. We set now
\begin{equation}\label{2.7}
\varphi_{\bfm}(\bfr,\alpha)=\varphi_{\jmath}(\bfr,\alpha)\equiv
\phi(\bfr-\sum_{l=1}^3m_l\bfa_l,\alpha).
\end{equation}
Function $\varphi_{\bfm}(\bfr,\alpha)$ results from the translation of
$\phi(\bfr)$ by $\sum_{l=1}^3m_l\bfa_l$.  It differs from zero only within
primitive cell $V_{\bfm}$ and is there normalized. Moreover the
$\varphi_{\bfm}(\bfr,\alpha)$s are orthonormal, \ie
\begin{equation}\label{2.7a}
\int_V \varphi_{j}(\bfr,\alpha)\varphi_{j'}
(\bfr,\alpha)dv =\delta_{j,j'},\quad j,j'=1,\ldots,N.
\end{equation}
A strongly localized quantum crystalline (SLQC) wave-function
is defined as the determinantal
function defined by Eq.s~(\ref{1.6}), (\ref{2.7}) and (\ref{2.7a}).
We assume that the normalized wave function of the $N$ polarized electrons
is a SLQC wave-function that, by the previous definitions,  is normalized
within $V$. With such a function, the electron number density $n(\bfr)$ turns
out to be periodic and equal to
\begin{equation}\label{2.7b}
n(\bfr)=\sum_{\bfm}\phi^2(\bfr +\sum_{i=1}^3 m_i\bfa_i,\alpha).
\end{equation}
As already anticipated, our task now is to evaluate the
expectation value $E(N,V,\alpha)$ of $\hH$ over a SLQC  wave-function and
to choose $\alpha$ so as to make $E(N,V,\alpha)$ as small as possible
for each $r_s$ value. After putting
\begin{equation}\label{2.8}
\epsilon(r_s,\alpha)\equiv E(N,V,\alpha)/N \equiv
\langle \psi\vert \hH\vert\psi\rangle /N,
\end{equation}
from Eq.s~(\ref{1.1}), (\ref{1.6}) and (\ref{2.7}) one gets
\begin{equation}\label{2.9}
\epsilon(r_s,\alpha) =   h_{b-b} + h_{e-b}+\langle \hT\rangle +
\langle \hV\rangle
\end{equation}
with
\begin{equation}\label{2.10a}
 h_{b-b}=\frac{e^2N}{2V^2}\int_V dv\int_V dv'
\frac{e^ {-\mu\vert\bfr - \bfr'\vert}} {\vert\bfr - \bfr'\vert},
\end{equation}
\begin{equation}\label{2.10c}
h_{e-b} =-\frac{e^2}{V}\sum_{j=1}^N\int_{V_0} dv\int_V dv'
\frac{{\phi}^2(\bfr,\alpha)e^ {-\mu\vert\bfr  - \bfr' +
\sum_{i=1}^3 m_i(j)\bfa_i\vert}}
{\vert\bfr - \bfr' +\sum_{i=1}^3 m_i(j)\bfa_i\vert},
\end{equation}
\begin{equation}\label{2.10}
\langle \hT\rangle =\frac{1}{2m}\int_{\V0}\nabla\phi(\bfr,\alpha)\cdot
\nabla\phi(\bfr,\alpha)dv,
\end{equation}
and
\begin{equation}\label{2.11}
 \langle \hV\rangle =\frac{e^2}{2N}\sum_{1\le \imath\ne\jmath\le N}
\bigl[J_{\imath,\jmath}-K_{\imath,\jmath}\Bigr],
\end{equation}
where\cite{Kohanoff}
\begin{eqnarray}
J_{\imath,\jmath} & \equiv & \int_V dv\int_V dv'
{\varphi_{\imath}}^2(\bfr,\alpha){\varphi_{\jmath}}^2(\bfr',\alpha)
\frac{e^{-\mu\vert \bfr-\bfr'\vert}}
{\vert \bfr-\bfr'\vert},\label{2.12}\\
K_{\imath,\jmath} & \equiv &  \int_V dv\int_V dv'
{\varphi_{\imath}}(\bfr,\alpha){\varphi_{\imath}}(\bfr',\alpha)
{\varphi_{\jmath}}(\bfr,\alpha){\varphi_{\jmath}}(\bfr',\alpha)
\frac{e^{-\mu\vert \bfr-\bfr'\vert}}
{\vert \bfr-\bfr'\vert}.\label{2.13}
\end{eqnarray}
Contributions $h_{b-b}$ and $h_{e-b}$ respectively originate from the
first and second term  on the right hand side
(rhs) of Eq.~(\ref{1.1}). Property (\ref{2.7})
has been used to derive Eq.s ~(\ref{2.10c}) and (\ref{2.10}).
Quantities $J_{\imath,\jmath}$ and $K_{\imath,\jmath}$ represent
the direct and the exchange contributions, respectively. The exchange
contributions originate from the antisymmetry of the wave-function.
In our case it turns out that the $K_{\imath,\jmath}$s with
$\imath\ne\jmath$, as the sum present in Eq.~(\ref{2.11}) requires,
are equal to zero owing to the fact that
${\varphi_{\imath}}(\bfr,\alpha)$ and ${\varphi_{\jmath}}(\bfr,\alpha)$
have supports with a void intersection. Thus, despite the fact that
the considered wave-function is antisymmetric,  SLQCS
are characterized by vanishing exchange contributions.
In other words,  SLQCSs are defined by the property that the
associated Hartree and Hartree-Fock equations yield the same final
results. In fact,  starting from Eq.~({1.7}), one easily
checks that  the resulting expectation of $\hH$ again is the sum of
analytical expressions (\ref{2.10}) and (\ref{2.18}). \ac
Using Eq.s~(\ref{2.7}) and (\ref{2.12}), the sum involving the
$J_{\imath,\jmath}$s becomes
\begin{equation}\nonumber
\sum_{1\le \imath\ne\jmath\le N}J_{\imath,\jmath}=
\sum_{\bfm\ne\bfm'}\int_{\V0}dv\int_{\V0}dv'\phi^2(\bfr,\alpha)
\phi^2(\bfr ',\alpha)
\frac{e^{-\mu\vert \bfr-\bfr'+\sum_{l=1}(m_l-{m_l}')\bfa_l\vert}}
{\vert \bfr-\bfr'+\sum_{l=1}(m_l-{m_l}')\bfa_l\vert}.
\end{equation}
With the change $(m_l-m_l')\to m_l$ and letting $N$ go to infinity one gets
\begin{equation}\label{2.15}
\sum_{1\le \imath\ne\jmath\le N}J_{\imath,\jmath}\to N{\sum_{\bfm}}'
\int_{\V0}dv\int_{\V0}dv'\phi^2(\bfr,\alpha)\phi^2(\bfr ',\alpha)
\frac{e^{-\mu\vert \bfr-\bfr'+\sum_{l=1}m_l\bfa_l\vert}}
{\vert \bfr-\bfr'+\sum_{l=1}m_l\bfa_l\vert},
\end{equation}
where $\bfm$ runs over all the points of the $\cZ^3$ lattice excluding the
origin. The last restriction is signaled by the prime.
Recalling that $V=\cup_{j=1}^N V_{\bfm(j)}$, contribution $h_{e-b}$  can be
converted into a double sum, \ie
\begin{equation}\nonumber
h_{e-b} =-\frac{e^2}{V}\sum_{\bfm,\bfm'}\int_{V_0} dv\int_{V_0} dv'
\frac{{\phi}^2(\bfr,\alpha)e^ {-\mu\vert\bfr  - \bfr' +
\sum_{i=1}^3 (m_i-{m'}_i)\bfa_i\vert}}
{\vert\bfr - \bfr' +\sum_{i=1}^3 (m_i-{m'}_i)\bfa_i\vert}.
\end{equation}
In the thermodynamic limit one finds that
\begin{equation}\label{2.16}
h_{e-b} =-e^2 n\sum_{\bfm}\int_{V_0} dv\int_{V_0} dv'
\frac{{\phi}^2(\bfr,\alpha)e^ {-\mu\vert\bfr  - \bfr' +
\sum_{i=1}^3 m_i\bfa_i\vert}}
{\vert\bfr - \bfr' +\sum_{i=1}^3 m_i\bfa_i\vert}.
\end{equation}
In a similar way one shows that
\begin{equation}\label{2.17}
h_{b
-b} =\frac{e^2 n^2}{2}\sum_{\bfm}\int_{V_0} dv
\int_{V_0} dv'\frac{e^ {-\mu\vert\bfr  - \bfr' +
\sum_{i=1}^3 m_i\bfa_i\vert}}
{\vert\bfr - \bfr' +\sum_{i=1}^3 m_i\bfa_i\vert}.
\end{equation}
Collecting the above results and taking the limit
$\mu\to\infty$ one obtains the contribution of the
total Coulombic interaction to the quantum energy
per particle, namely 
\begin{eqnarray}
\ &&h_{b-b} + h_{e-b}+\langle \hV\rangle=\frac{e^2}{2}\biggl[-
\int_{V_0} dv\int_{V_0} dv'\frac{2n\phi^2(\bfr,\alpha)-n^2}
{\vert\bfr - \bfr'\vert}\nonumber\\
\ & &+{\sum_{\bfm}}'\int_{V_0} dv\int_{V_0} dv'
\frac{\phi^2(\bfr,\alpha)\phi^2(\bfr',\alpha) - 2n\phi^2(\bfr,
\alpha)+n^2}
{\vert\bfr - \bfr' +\sum_{i=1}^3 m_i\bfa_i\vert}
\biggr].\label{2.18}
\end{eqnarray}
We make now three remarks. First, if one assumes that
$\phi^2(\bfr,\alpha)$ is a Gaussian function, the rhs of
Eq.~(\ref{2.18}) coincides with the expression reported
by Ewald in his calculation of the Coulombic energy
per particle for a classical crystal (see \eg\
Ref.s~\cite{Slater,AshMer}). However, the introduction of the
Gaussian function in Ewald's procedure was only a trick to
make the convergence faster, while in the SLQCS procedure
$\phi^2(\bfr,\alpha)$ is the quantum probability density of
finding an electron at position $\bfr$. Second, the series
present in Eq.~(\ref{2.18}) is convergent. In fact, setting
$A(\bfm)\hnu(\bfm)\equiv\sum_{l=1}m_l\bfa_l $ where
$\hnu(\bfm)$  is a unit vector, the expansion of the
denominator, present in Eq.~(\ref{2.18}),
at large $A(\bfm)$ values yields
\begin{eqnarray}
\frac{1}{\vert \bfr-\bfr'+A(\bfm)\hnu(\bfm)\vert}&\approx &\frac{1}{A(\bfm)}
\Bigl[ 1  - \frac{(\bfr-\bfr')\cdot\hnu(\bfm)}{A(\bfm)}+\nonumber\\
\quad & & +\frac{3(\bfr-\bfr')\cdot\hnu(\bfm)-(\bfr-\bfr')^2}{2A^2(\bfm)}
+\ldots\Bigr].\label{2.20}
\end{eqnarray}
The volume integrals of the first term on the rhs is equal to zero. The
remaining two terms also yield vanishing contributions provided the sum
over $\bfm$ is first performed over all the $\bfm$s that have different
directions and equal modulus and, subsequently, over the different
$|\bfm|$ values. In this way,  as $A(\bfm)$ becomes very large,  the first
of the resulting sums amounts to performing an angular average over the
directions of $\hnu(\bfm)$. It is easily checked that the $\hnu$ dependence
of the numerators present in Eq.~(\ref{2.20}) is such that their angular
averages  vanish for the sc, fcc and bcc cases. One
concludes  that the terms of series (\ref{2.18}) decrease as
$A^{-4}(\bfm)$ at large $\bfm$s, and the series convergence is proved.
Third, owing to condition (\ref{2.4}), the limit of Eq.~(\ref{2.18}) as
$\alpha\to\infty$ is the electrostatic energy of a lattice of point-like
charges plus a uniform neutralizing charge within each cell (the cubic
symmetry being specified by $V_0$ and the $\bfa_i$s).
Hence, the limit value of  the rhs of (\ref{2.18}) for $\alpha\to\infty$
is equal to ${e^2 M_{dl,\sigma}}/{2a_0 r_s}$ where the
$M_{dl,\sigma}$s are the Madelung constants defined below Eq.~(\ref{1.3}).
This remark shows that the
SLQC approximation yields the classical crystal energy in the infinite
dilution limit because later, after illustrating the results shown in
Fig.2, it will appear clear that the limit $r_s\to\infty$ implies that
$\alpha\to\infty$. \ac
Before concluding the section, we further elaborate the above
expressions in terms of new integration variable $\vxi$ and dimensionless
function $\phi_0(\vxi,\alpha)$.  The new quantities are respectively
defined by
$$\bfr=\sum_{l=1}^3\xi_l\bfa_l/2,$$
and
\begin{equation}\label{2.21}
\phi_0(\vxi,\alpha)\equiv (8/v_0)^{-1/2}\phi(\bfr,\alpha)=
(8/v_0)^{-1/2}\phi(\sum_{l=1}^3\xi_l\bfa_l/2,\alpha).
\end{equation}
Recalling Eq.~(\ref{2.1}), the tip of vector $\vxi$ is confined to the
cubic cell $C_0$ of edge 2 centred at the origin of an orthogonal
Cartesian frame and defined as
$$C_0\equiv\{\vxi| -1\le \xi_j\le1,\quad j=1,2,3\}.$$
$\phi_0(\vxi,\alpha)$ is an even function and, due to
Eq.~(\ref{2.3}), it obeys to
\begin{equation}\label{2.22}
\int_{\C0}{\phi_0}^2(\vxi,\alpha)d^3\vxi = 1.
\end{equation}
Substituting $(v_0/8)^{1/2}\phi_0(\vxi,\alpha)$ for $\phi(\bfr,\alpha)$
in Eq.s~(\ref{2.10}) and (\ref{2.18}) and passing to new integration variable
$\vxi$,
one finds that
\begin{equation}\label{2.23}
\epsilon_{\sigma}(r_s,\alpha)= \frac{e^2}{2a_0}\biggl[
\frac{\tau_{\sigma}(\alpha)}{{r_s}^2} +
\frac{\upsilon_{{\sigma}}(\alpha)}{r_s}\biggr]
\end{equation}
where the dimensionless quantities $\tau_{\sigma}(\alpha)$ and
$\upsilon_{{\sigma}}(\alpha)$  are
\begin{eqnarray}
\tau_{\sigma}(\alpha) &\equiv &\kappa_{\sigma}\sum_{j,l=1}^3
\int_{\C0}\cA^{\sigma}_{j,l}
(\partial_j\phi_0(\vxi,\alpha))(\partial_l\phi_0(\vxi,\alpha))
d^3\vxi,\label{2.24}\\
 \upsilon_{\sigma}(\alpha)&\equiv& \chi_{\sigma}\biggl[
{\sum_{\bfm}}'\int_{\C0}d^3\vxi\int_{\C0}d^3\vxi'
\frac{{\phi_0}^2(\vxi,\alpha){\phi_0}^2(\vxi',\alpha) -
{\phi_0}^2(\vxi,\alpha)/4 +1/64}
{d_{\sigma}(\vxi - \vxi'+2\bfm)}\nonumber \\
 \quad & & \quad\quad - \int_{\C0}d^3\vxi\int_{\C0}d^3\vxi'
\frac{{\phi_0}^2(\vxi,\alpha)/4 -1/64}
{d_{\sigma}(\vxi-\vxi')}\biggr]. \label{2.25}
\end{eqnarray}
In these two equations, similarly to $M_{dl,\sigma}$'s definition,
index $\sigma$ ranges over $\{1,2,3\}$ respectively associated
to  lattices sc, bcc and fcc.  Moreover, the partial derivatives
present in Eq.~(\ref{2.24}) refer to variable $\vxi$ while the
remaining symbols are defined as follows:
\begin{eqnarray}
& & \kappa_1=\Bigl(\frac{6}{\pi}\Bigr)^{2/3},\  \
{\cA^{(1)}}_{j,l}=\delta_{j,l}, \ \
\chi_{1}= \Bigl(\frac{6}{\pi}\Bigr)^{1/3},\nonumber \\
& & \ \ d_{1}(\vxi)\equiv\Bigl[\sum_{j=1}^3
\xi_j^2\Bigr]^{1/2} \label{2.27}
\end{eqnarray}
for the sc case,
\begin{eqnarray}
& & \kappa_2=\Bigl(\frac{3}{\pi}\Bigr)^{2/3},\ \
{\cA^{(2)}}_{j,j}=2,\ {\rm and}\ {\cA^{(2)}}_{j,l}=1\ {\rm if}\ j\ne l,\ \
\chi_{2}= \Bigl(\frac{8}{\sqrt{3}\pi}\Bigr)^{1/3},\nonumber \\
& &d_{2}(\vxi)\equiv\Bigl[\sum_{j=1}^3
{\xi_j}^2 -\frac{2}{3}\sum_{1\le j<l\le 3}\xi_j
\xi_l \Bigr]^{1/2}\label{2.28}
\end{eqnarray}
for the bcc case, and
\begin{eqnarray}
& &\kappa_3=\Bigl(\frac{3}{2\pi}\Bigr)^{2/3},\ \
{\cA^{(3)}}_{j,j}=3 \ {\rm and}\ {\cA^{(3)}}_{j,l}=-1\ {\rm if}\ j\ne l,\ \
\chi_{3}= \Bigl(\frac{3\sqrt{2}}{\pi}\Bigr)^{1/3},\nonumber\\
& & d_{3}(\vxi)\equiv\Bigl[\sum_{j=1}^3
{\xi_j}^2   +\sum_{1\le j<l\le 3}\xi_j \xi_l \Bigr]^{1/2} \label{2.29}
\end{eqnarray}
for the fcc case.\ac
Numerically it is more convenient to express $\upsilon_{{\sigma}}(\alpha)$
in terms of Fourier transforms (FT) as we already showed in a first
presentation\cite{Cicca} of the SLCQS approach based on quantum field 
theory but restricted to the only sc case.
After denoting the FT of ${\phi_0}^2(\vxi,\alpha)$ by $\tphi2(\bfq,\alpha)$,
\ie
\begin{equation}\label{FT}
\tphi2(\bfq,\alpha)  =
\int d^3\vxi e^{-i\bfq\cdot\vxi}\
{\phi^2}_0(\vxi,\alpha),
\end{equation}
in appendix A we  show that $\upsilon_{{\sigma}}(\alpha)$
can be written as
\begin{equation}\label{2.30}
\upsilon_{\sigma}(\alpha) =  \upsilon_{P,\sigma}(\alpha) +
\upsilon_{N,\sigma}(\alpha),
\end{equation}
with
\begin{eqnarray}
\upsilon_{P,\sigma}(\alpha) &\equiv& {{\chi_{\sigma}'}}
{\sum_{\bfm}}'
\frac{(\tphi2(\pi\bfm,\alpha))^2}{{\td_{\sigma}}(\bfm)},\label{2.31}\\
\upsilon_{N,\sigma}(\alpha) &\equiv&-\chi_{\sigma}\int_{C_0} d^3\vxi
\int_{C_0} d^3\vxi'
\frac{{\phi_0}^2(\vxi,\alpha){\phi_0}^2(\vxi',\alpha)}
{{d_{\sigma}}(\vxi-\vxi')},\label{2.32}\\
\td_1(\bfq)&\equiv& {q_1}^2 +{q_2}^2 +{q_3}^2,\label{2.33}\\
\td_2(\bfq)&\equiv& \bfq\cdot\bfq + {q_1}q_2 +{q_2}q_3 +{q_3}q_1,
\label{2.34}\\
\td_3(\bfq)&\equiv& \bfq\cdot\bfq -
2({q_1}q_2 +{q_2}q_3 +{q_3}q_1)/3,\label{2.35}
\end{eqnarray}
and
\begin{equation}\label{2.35bis}
{\chi_1}'\equiv \chi_1/2\pi,\quad
{\chi_2}'\equiv \sqrt{3}\chi_2/4\pi \quad {\rm and}\quad
{\chi_3}'\equiv2^{2/3}\chi_3/6\pi.
\end{equation}
After choosing a particular real even function $\phi_0(\vxi,\alpha)$,
which obeys Eq.~(\ref{2.22}) and vanishes with its first and second partial
$\xi$-derivatives on the boundary of $\C0$, the best  SLQC wave-function is
obtained as follows. First, one evaluates quantities $\tau_{\sigma}(\alpha)$,
$\upsilon_{P,\sigma}(\alpha)$ and $\upsilon_{N,\sigma}(\alpha)$ by
Eq.s~(\ref{2.24}), (\ref{2.31}) and (\ref{2.32})
over a grid of $\alpha$ values: $\alpha_1,\ldots,\alpha_L$. Second, chosen an
$r_s$ value, by Eq.s~(\ref{2.23}), (\ref{2.24}), (\ref{2.31}) and (\ref{2.32})
one obtains the set
of values $\epsilon_{\sigma}(r_s,\alpha_i)$ with $i=1,\ldots,L$. The smallest
of these values will correspond to a particular $i$ denoted by ${\bar i}$.
Then, $\epsilon_{\sigma}(r_s,\alpha_{\bar i})$ approximates the energy of the
fundamental state for the considered SLQC wave-function with the considered
crystalline symmetry, while $\alpha_{\bar i}$ represents the value of
$\alpha_{\sigma}(r_s)$. Finally, the value of $\sigma$ that yields the
smallest energy at a fixed $r_s$ value determines the crystalline symmetry
of the jellium at the considered density.

\section{Numerical results}
We report now a numerical illustration of the procedure just described.
To this aim we define function $\phi_0(\vxi,\alpha)$ as follows
\begin{equation}\label{3.1}
\phi_0(\vxi,\alpha)\equiv \prod_{j=1}^3 G(\xi_j,\alpha)
\end{equation}
with
\begin{eqnarray}
G(\xi,\alpha)&\equiv& C(\alpha)e^{-\alpha\xi^2/(1-\xi^2)},\label{3.2}\\
C(\alpha)&\equiv&\Bigl(\sqrt{\pi}\Psi({1}/{2},0;2\alpha)\Bigr)^{-1/2},
\label{3.3}
\end{eqnarray}
where $\Psi({1}/{2},0;2\alpha)$ is a particular value
of the confluent Hypergeometric function $\Psi(a,c;z)$
defined in \S 6.5 of Ref.\cite{ErdMaObTr}.
The reported $C(\alpha)$ expression ensures that
$\phi_0(\vxi,\alpha)$ obeys condition (\ref{2.22}).
Besides, this function as well as  all its partial derivatives
vanish as $\vxi$ approaches the border of $\C0$.
The factorized expression of $\phi_0(\vxi,\alpha)$ further simplifies
Eq.s~(\ref{2.24}), (\ref{FT}), (\ref{2.31}) and (\ref{2.32}). In fact,
as shown in appendix B, $\tau_{\sigma}(\alpha)$ becomes
\begin{equation}\label{3.4}
 \tau_{\sigma}(\alpha) = {\kappa'}_{\sigma}\tau(\alpha)\equiv
{\kappa'}_{\sigma}\,\Bigl[2\,\int_0^1
[\partial_{\xi}G(\xi,\alpha)]^2d\xi
\Bigr]\end{equation}
with ${\kappa'}_1\equiv 3{\kappa}_1$, ${\kappa'}_2\equiv 6{\kappa}_2$,
${\kappa'}_3\equiv 9{\kappa}_3$ and
\begin{equation}\label{3.5}
\tau(\alpha)=  \frac{\Psi(-\frac{3}{2},\,-2;\, 2\alpha)}
                     {4\alpha\,\Psi(\frac{1}{2},\,0\,;\,2\alpha)}.
\end{equation}
$\tphi2(\bfq,\alpha)$ becomes
\begin{equation}\label{3.6}
\tphi2(\bfq,\alpha) = \prod_{j=1}^3 \tG2(q_j,\alpha)
\end{equation}
with
\begin{equation}\label{3.7}
\tG2(q,\alpha)= \int_{-1}^1d\xi e^{-iq\xi}G^2(\xi,\alpha)=
2\int_{0}^1d\xi \cos(q\xi)G^2(\xi,\alpha).
\end{equation}
Besides, after putting for $0\le\eta\le 2$
\begin{equation}\label{3.8}
\Gamma(-\eta,\alpha)=\Gamma(\eta,\alpha)\equiv \int_{-1}^{1-\eta} 
G^2(\eta+\xi',\alpha)G^2(\xi',\alpha)d\xi',
\end{equation}
%for $\upsilon_{P,\sigma}(\alpha)$ and $\upsilon_{N,\sigma}(\alpha)$
Eq.s~(\ref{2.31}) and (\ref{2.32}) respectively become 
\begin{eqnarray}
 \upsilon_{P,\sigma}(\alpha)&=& {{\chi_{\sigma}}'}
{\sum_{\bfm}}'  \frac{\prod_{j=1}^3 (\tG2(\pi m_j,\alpha))^2}
{{\td_{\sigma}}(\bfm)},\label{3.9}\\
 \upsilon_{N,\sigma}(\alpha)& = & - \chi_{\sigma}
\int_{2\C0}d^3{\vec\eta}
\frac{\prod_{l=a}^3\Gamma(\eta_l,\alpha)}{d_{\sigma}({\vec \eta})},
\label{3.10}
\end{eqnarray}
where $2\C0$ denotes the cubic cell $\{{\vec\eta}\,
\bigl\vert\,  -2\le \eta_i\le 2,\ i=1,2,3\}$, and with the
$\td_{\sigma}(\bfm)$s and $d_{\sigma}({\vec \eta})$s defined by
Eqs.~(\ref{2.33})-(\ref{2.35}) and (\ref{2.27})-(\ref{2.29}).\ac
In this way, we must numerically evaluate the 1D integral present on the
rhs of (\ref{3.4}) to determine $\tau_{\sigma}(\alpha)$,
the 1D FT defined by Eq.~(\ref{3.7}) at a set of values $q=\pi m$ with
$m=0,\ldots,M$ to subsequently evaluate series (\ref{3.9}) truncated at $M$,
and, finally, $\Gamma(\eta,\alpha)$ at a set of points
$\eta_1,\ldots,\eta_{M_I}$ lying within the interval [0,2] to evaluate
 3D integral (\ref{3.10}). (By so doing, we use the eveness of 
$\Gamma(\eta,\alpha)$ and $\tG2(q,\alpha)$ with respect to  $\eta$ and $q$.) 
The calculations must performed over a grid of $\alpha$
values: $0<\alpha_1<\ldots<\alpha_{L}$. After numerically determining
$\tau_{\sigma}(\alpha_i)$ by Eq.~(\ref{3.4}), $\upsilon_{P,\sigma}(\alpha_i)$
by Eq.~(\ref{3.9}) and $\upsilon_{N,\sigma}(\alpha_i)$ by Eq.~(\ref{3.10}),
we look for the minimum value of
\begin{equation}\label{3.11}
\epsilon_{\sigma}(r_s,\alpha_i) = \frac{\tau_{\sigma}(\alpha_i)}{{r_s}^2} +
\frac {\upsilon_{P,\sigma}(\alpha_i)+\upsilon_{N,\sigma}(\alpha_i)}{r_s}
\end{equation}
at a fixed $r_s$ value as $i$ ranges from 1 to $L$. Then, as already
reported at the end of the previous section, if the minimum occurs at
$i={\bar i}$, the SLQC fundamental state corresponds to Eq.~(\ref{1.6})
with $\alpha=\alpha_{\bar i}$ and the corresponding energy per particle,
in Rydberg units, is $\epsilon_{\sigma}(r_s,\alpha_{\bar i})$. Changing
the $r_s$ value, ${\bar i}$ will also change and the set of the
$\alpha_{\bar i}$ values numerically determines $\alpha_{\sigma}(r_s)$, \ie\
the way $\alpha_{\sigma}$ changes with $r_s$ is determined by the set of
values $\alpha_{\bar i}$ relevant to the considered lattice symmetry.\ac
Before illustrating our numerical results, we give some details about
numerical computations.  We considered an
 $\alpha$ grid that spans the interval [0.01, 500] at integer multiples of
0.05 up to $\alpha=0.5$, of 0.1 up to $\alpha=2$, of 1 up to $\alpha=20$,
of 10 up to $\alpha=100$ and of 100 up to $\alpha=500$. The numerical accuracy
of the results, depending on the values of $M$, $M_I$
and the number of points used to evaluate quantities
$\tau(\alpha)$, $\tG2(q,\alpha)$, $\Gamma(\eta,\alpha)$ and
$\upsilon_{\sigma}(\alpha)$,  was tested at $\alpha=0.01$ and 100.
The first six significant digits of $\tau(\alpha)$ do not change
passing from $10^8$ to $10^9$ points. The first 5 digits of the FTs
evaluated with $10^6$  or $10^7$ points do not differ and agree with
their asymptotic formula given by Eq.~(96) of \cite{Cicca} [note that
here the
correct argument of the sine function is $(2+\sqrt{2q\alpha}-\pi/8$)].
The first five digits of the resulting $\upsilon_{P,\sigma}(\alpha)$
values, obtained truncating the series at $M=50$ or $M=100$, also coincide.
The calculation of $\upsilon_{N,\sigma}(\alpha)$ requires more care.
The evaluation of $\Gamma(\eta)$ with integration steps of $10^{-5}$
or $10^{-6}$ leaves the first five digits unchanged. However, the evaluation
of the remaining 3D integral over $\vec\eta$ cannot be done with
an integration step $\Delta\eta$  smaller than $10^{-4}$ because
the required CPU time becomes very large.  Thus we made three runs with
$\Delta \eta$ respectively equal to 1/500, 1/1000 and 1/1500. The first
four digits do not change even at large $\alpha$s, \ie\ $\alpha\ge 100$,
which is the critical region. The ratio of the resulting errors, defined as
(second -fist)run divided by (third - first)run, was found fairly equal to
3/2. In this way we numerically extrapolated the results to $\Delta\eta\to 0$.
The resulting $\upsilon_{N,\sigma}(\alpha)$ values ought to have five
correct digits. $\upsilon_{P,\sigma}(\alpha)$ and $\upsilon_{N,\sigma}
(\alpha)$ always have opposite signs. At large $\alpha$s, they have the
first two digits equal. Hence, the final $\upsilon_{\sigma}(\alpha)$ values
have the first three digits exact for all the considered values of $\alpha$.\ac
We pass now to illustrate our results. These are shown in Fig.s~1a, 1b and 2.
Fig.s~1a and 1b respectively show the energy per particle (in Rydberg units)
{\em versus} $r_s$, in the high density region, and {\em vs} $Lg_{10}(r_s)$
in the low density one. In fact, the continuous and dotted curves refer to 
Bloch's
expression (\ref{1.3}) and Carr's expression (\ref{1.5}), respectively. The
long-dash curves refer to the optimized SLQC wave-function with symmetry bcc,
and the short-dash curves to the simple cubic optimized SLQC wave-function.
The full triangles and circles are the values respectively obtained for the
fluid and bcc crystalline phases by QMC calculations\cite{CeperAl,OrtiBa,Zong,
DruRaTrToNe}. (We have not reported the curve relevant to the fcc SLQCS
solution in order to not overcrowd the figure. It lies close to the bcc 
solution.) The short-dash curve crosses the continuous one at $r_s=28$ and
for greater $r_s$ values it lies below the continuous curve. On the basis of
the Ritz-Rayleigh principle one concludes that the SLQCS with the sc symmetry
is closer to the true fundamental state than $|F_p\rangle$.
In other words, as the density decreases, the jellium passes from the polarized
fluid phase to the sc crystalline one at $r_s=28$. The figures also show that
long-dash curve passes from above to below the continuum one at $r_s=38$.
This means that in the region $28 < r_s < 38$, the sc phase is more stable
than the fluid which in turn is more stable than the bcc. In the region
$r_s>38$, the bcc phase is more stable than the fluid and is less
stable than the sc up to $r_s=500$. The figures make also evident a property
that appears to have been overlooked by most textbooks: Carr's approximation
appears to be surprisingly accurate throughout the full density range since
it  fairly agrees with QMC results even at high densities.\ac
We pass now to the illustration of figure 2.
\begin{figure}
\begin{center}
%{{
\includegraphics[width=0.8\textwidth]{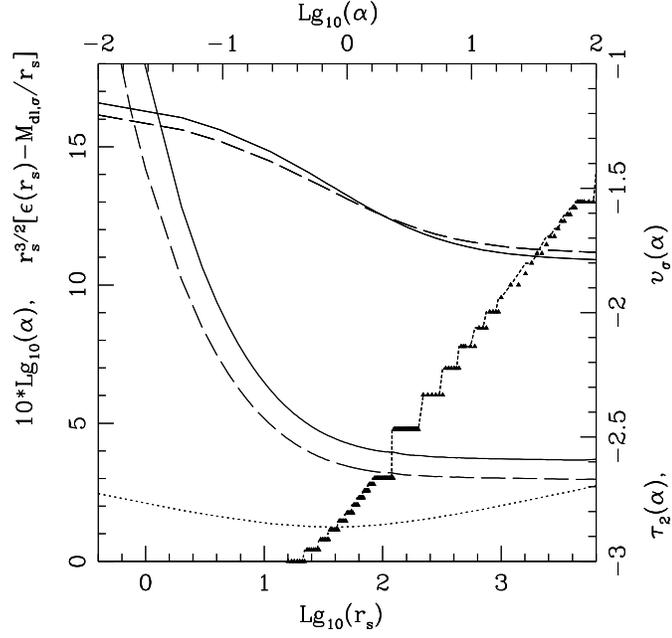}%\par
%}
\end{center}
\vskip -3.truecm
{\caption {\label{fig2} The dotted step-line and the full triangles are the
			       plots of
$10\times  Lg_{10}(\alpha_{\sigma})$ {\em vs.} $Lg_{10}(r_s)$ for the sc and
bcc case; the continuous and the long-dash curves (convex and monotonically
decreasing) those of $r_s^{3/2}[\epsilon_{\sigma}(r_s)+
M_{dl,\sigma}/r_s]$ {\em vs.} $Lg_{10}(r_s)$ for the bcc ($\sigma=2$)
and sc case ($\sigma=1$) [the corresponding vertical scale is on the
left and the energy units are ryd];  the continuum and the
long-dash sigmoidal curves those of $\upsilon_2(\alpha)$ and
$\upsilon_1(\alpha)$ {\em vs.} $Lg_{10}(\alpha)$ (the relevant horizontal
and vertical scales are the top and the right ones); finally, the
parabolic dotted curve that of $Lg_{10}(\tau_2(\alpha))$ {\em vs.}
$Lg_{10}(\alpha)$ (its vertical scale is the left one).
}}
\end{figure}
The full triangles and the close dotted line respectively represent
$\alpha_2(r_s)$ and $\alpha_1(r_s)$ on a log-log scale. The apparent
step behaviour is an artifact of the chosen $\alpha$-grid whose values
are not equally spaced. One should note, at large $Lg_{10}(r_s)$s, the 
approximate linear behaviour of $Lg_{10}(\alpha)$ with a slope 
corresponding to have $\alpha_{\sigma}\propto {r_s}^{1/2}$. The
continuous and the long-dash curve reported on the top of the
figure respectively plot $\upsilon_{2}(\alpha)$
and $\upsilon_1(\alpha)$ {\em vs.} $Lg_{10}(\alpha)$. 
These curves can also be considered as the plots of 
$\upsilon_{\sigma}(\alpha(r_s))$ {\em vs.} $r_s$ for 
the noted property that $\alpha_{\sigma}\propto {r_s}^{1/2}$.   In this way,
the figure makes it evident that the SLQCS $\upsilon_{\sigma}(r_s)$s 
approach the relevant Madelung values as $r_s\to\infty$. The 
remaining two monotonically decreasing convex curves  are the
plots of  $r_s^{3/2}[\epsilon_{\sigma}(r_s)+M_{dl,\sigma}/r_s]$ 
{em vs.} $r_s$  for the bcc (continuous curve) and the sc (dotted) symmetry.
The constant behaviour observed at large $r_s$ values indicates
that
\begin{equation}\label{3.12}
\epsilon_{\sigma}(r_s)\approx -\frac{M_{dl,\sigma}}{r_s} +
\frac{{\cal C}_{\sigma}}{{r_s}^{3/2}}+o\big(\frac{1}{{r_s}^{3/2}}\bigr),
\end{equation}
the ${\cal C}_{\sigma}$s being
appropriate constants that from the figure appear to be equal to 3.6 and 2.9
for $\sigma=2$ and 1, respectively. In appendix B we show that
the SLQCS procedure always yields an energy per particle that asymptotically
behaves as reported in Eq.~(\ref{3.12}) at large $r_s$. It is stressed that
this behaviour coincides with that of Eq.s~(\ref{1.4}) and (\ref{1.5}), even
though the numerical coefficients of the $r_s^{-3/2}$ contribution are
different. In this respect, we recall that the crystalline HF solutions
investigated in Ref.\cite{Trail} also behave as in Eq.~(\ref{3.12}) and from
Fig.4 of this paper it appears that ${\cal C}_2\approx 3$. We once more
underline that contribution  $-{M_{dl,2}}/{r_s}$ comes naturally out by
the HF and the SLQCS procedure [that is equivalent to a Hartree  or a HF
equation as  explained below Eq.~(\ref{2.13})], while in 
 Wigner's and Carr's derivation of Eq.s~(\ref{1.4}) and
(\ref{1.5}) it was put there because it is 
the energy of the 'static' unperturbed Hamiltonian.\ac
Finally, the dotted parabolic curve  plots
$Lg_{10}(\tau_2(\alpha))$ {\em vs} $Lg_{10}(\alpha)$. It shows a  linear
behaviour in the outermost $\alpha$ range that follows from Eq.~(\ref{3.5})
as we explicitly show in appendix B.

\section{Conclusion}
As Wigner first pointed out, an overall neutral one component plasma
of electrons shows, at $T=0^0K$, the interesting feature of being in a
crystalline phase as the particle number density becomes smaller than
a particular value. This feature can nowadays be considered well assessed
because it was confirmed by HF and QMC calculations (though the transition
density is not accurately known yet). The fact that quantum Coulombic
crystals exist at very high dilution, on physical grounds, has the
important consequence that their behaviours must approach those of the
corresponding classic Coulombic crystals. In fact, in the infinite dilution
limit, the inter-electron mean distance becomes infinitely large with
the consequence that: the overlapping among the the wave-functions of
different electrons is expected to vanish and electrons become
distinguishable in the sense that one can speak of an electron that
occupies a well definite region/cell of the crystal.  The SLQCS approach
exactly captures the last two features. In fact, the SLQCS is completely
antisymmetric but the support properties of the involved wave-functions
make the fully antisymmetric wave-function equivalent to the Hartree one.
Consequently, the electrons behave as particles each of them being
confined to a single cell of the crystal. The determination of the
{\em best} SLQCS, equivalent to solve the corresponding Hartree or 
Hartree-Fock equation,  proceeds by a simple variational procedure. In
fact, we worked out the expression, in an integro-differential
form, of the energy per particle, given by Eq.~(\ref{2.23})
together with Eq.s~(\ref{2.24})-(\ref{2.35}), in terms of the basic
function $\phi(\bfr,\alpha)$ different from zero only within a
single primitive cell and depending  on a real parameter $\alpha$.
The solution was obtained by looking for the minimum of
the energy with respect to $\alpha$ at each fixed $r_s$ value.
The results illustrated in the three figures show that, at large
$r_s$s, the leading term of the SLQCS energy per particle coincides
with that of Eq.s(\ref{1.4}) and (\ref{1.5}) with an important
difference: Wigner and Carr took this contribution from the classical
Coulombic crystal's value while the SLQCS procedure directly produces
this term. Moreover, as shown in appendix B, this procedure also implies
that the next to the leading asymptotic term of $\epsilon_{\sigma}(r_s)$
behaves  as $C_{\sigma}/{r_s}^{3/2}$. The $r_s$ dependence coincides
with that of  Wigner's and Carr's formulae but the numerical values of
coefficient $C_{\sigma}$ are different. These values are also different
from the values obtained solving the HF equation\cite{Trail}. Hence,
in the very low density region, the differences among the Wigner/Carr,
the SLQCS and the HF  approximations of the jellium model set only into
at the level of the $O({r_s}^{-3/2})$ term.
We consider this result as the main conclusion of this analysis. The
implications of this conclusion are: Wigner or Carr formulae are
equally accurate in the far $r_s$ region though the dominating
contribution was not quantum-mechanically derived; the SLQCS approach
quantum-mechanically derives this contribution and therefore it
is there as accurate as the HF equation but of simpler application;
as the system becomes denser the numerical differences in coefficient
$C_{\sigma}$ make the approximations no longer equivalent. One expects
that overlapping effects be no longer negligible and that the SLQCS
approach drastically deteriorates in comparison to the HF one.

%%%%%\leftline{\bfquindc Appendices}
\appendix{}
\section{ Conversion of Eq.~(\ref{2.25}) in Eq.s~(\ref{2.31})
and (\ref{2.32})}
To write Eq.s~(\ref{2.25}) in a form more convenient for numerical
computations, we first introduce function $\Theta_0(\vxi)$,
equal to 1 if $\vxi$ lies within $C_0$ and equal to zero elsewhere.
After putting
\begin{equation}\label{a1}
\Psi(\vxi,\alpha)\equiv {\phi_0}^2(\vxi,\alpha) -\Theta_0(\vxi)/8,
\end{equation}
one easily proves that the quantity inside the square brackets on
the rhs of Eq.~(\ref{2.25}) can be written as
\begin{equation}\label{a2}
{\sum_{\bfm}}\int_{\C0}d^3\vxi\int_{\C0}d^3\vxi'
\frac{\Psi(\vxi,\alpha)\Psi(\vxi',\alpha)}
{d_{\sigma}(\vxi - \vxi' + 2\bfm)}
 - \int_{\C0}d^3\vxi\int_{\C0}d^3\vxi'
\frac{{\phi_0}^2(\vxi,\alpha){\phi_0}^2(\vxi',\alpha)}
{d_{\sigma}(\vxi - \vxi')},
\end{equation}
where the sum includes now the term with $\bfm={\bf 0}$. Since
$\phi_0(\vxi,\alpha)$ and $\Theta_0(\vxi)$ are equal to zero outside $C_0$,
integrals in Eq.~(\ref{a2}) can be extended to the full $R^3$. The FT
transform of
$\Psi(\vxi,\alpha)$ will be denoted by ${\tilde\Psi}(\bfq,\alpha)$. Both
${\phi_0}(\vxi,\alpha)$ and  $\Theta_0(\vxi)$ are real and even function of
$\vxi$. Then ${\tilde\Psi}(\bfq,\alpha)$ is an even function of $\bfq$. Using
the FTs, the generic term of the series can be written as
\begin{equation}\label{a3}
\frac{1}{(2\pi)^6}\int_{R^3}d^3\vxi\int_{R^3}d^3\vxi'
\frac{1}{d_{\sigma}(\vxi - \vxi' + 2\bfm)}\int d^3\bfq e^{i\bfq\cdot\vxi}\
{\tilde\Psi}(\bfq,\alpha) \int d^3\bfq'  e^{-i\bfq'\cdot\vxi'}
\ {\tilde\Psi}(\bfq',\alpha).
\end{equation}
The change of the integration variable
$\vxi \to {\vec\eta}=\vxi-\vxi'+2\bfm$ allows us to perform the $\vxi'$
integration so as to convert the previous expression into
\begin{equation}\nonumber
\frac{1}{(2\pi)^3}\int_{R^3}d^3{\vec \eta}\int d^3\bfq
\frac{1}{d_{\sigma}({\vec \eta})}e^{i\bfq\cdot({\vec\eta}-2\bfm)}
\ {\tilde\Psi}^2(\bfq,\alpha).
\end{equation}
The integral over ${\vec \eta}$ yields
\begin{equation}\nonumber
\frac{1}{(2\pi)^3}\int_{R^3}d^3{\vec \eta}
\frac{1}{d_{\sigma}({\vec \eta})}e^{i\bfq\cdot{\vec\eta}}=
\frac{4\pi}{(2\pi)^3}\frac{\omega_{\sigma}}{\td_{\sigma}(\bfq)}
\end{equation}
with
\begin{equation}\label{a3bis}
\omega_1\equiv 1,\quad\omega_2\equiv\sqrt{3}/2,\quad
\omega_3\equiv 2^{2/3}/3,
\end{equation}
and the $\td_{\sigma}(\bfq)$s defined by Eq.s~(\ref{2.33})-(\ref{2.35}).
In this way, the series present in Eq.~(\ref{a2}) converts in
\begin{equation}\label{a4}
\frac{4\pi\omega_{\sigma}}{(2\pi)^3}\int d^3\bfq
\frac{{\tilde\Psi}^2(\bfq,\alpha)}{\td_{\sigma}(\bfq)}
{\sum_{\bfm}}e^{-i2\bfm\cdot\bfq}.
 \end{equation}
Using the  mathematical identity\cite{Marder}
\begin{equation}\label{identit}
{\sum_{\bfm}}e^{i2\bfm\cdot\bfq}  =
\frac{(2\pi)^3}{8}\sum_{\bfm}\delta(\bfq - \pi\bfm)
\end{equation}
the integrals over $\bfq$ in Eq.~(\ref{a4}) are immediately
evaluated if $\bfm\ne{\bf 0}$, while the contribution relevant
to $\bfm={\bf 0}$ is equal to zero because the resulting integrand
${{\tilde\Psi}^2(\bfq,\alpha)}/{\td_{\sigma}(\bfq)}$ vanishes at
$\bfq={\bf 0}$. In fact, condition (\ref{2.22}) and the definition of
$\Theta_0(\vxi)$ imply that
$${\tilde\Psi}({\bf 0},\alpha) = \int_{C_0}d^3\vxi\ \bigl[
{\phi_0}^2(\vxi,\alpha)-\Theta_0(\vxi)/8\bigr] = 0.$$
Moreover the eveness and the reality  of $\phi_0(\vxi,\alpha)$ and
$\Theta_0(\vxi)$ ensure that
$${\tilde\Psi}(\bfq,\alpha)=\tphi2(\bfq,\alpha) - \tt0(\bfq)/8 \approx
O(\bfq\cdot\bfq)$$ at small $|\bfq|$. This property implies that
$$\frac{{\tilde\Psi}^2(\bfq,\alpha)}{\td_{\sigma}(\bfq)}
\approx O(|\bfq|^2)$$
and one concludes that no contribution to the sum over ${\bfm}$
arises from the term with $\bfm={\bf 0}$. Thus, the series present
in Eq.~(\ref{a2}) is equal to
\begin{equation}\label{a5}
\frac{\omega_{\sigma}}{2\pi}{\sum_{\bfm}}'\frac{{\tilde\Psi}^2(\pi\bfm,\alpha)}
{\td_{\sigma}(\bfm)}.
\end{equation}
A further simplification follows from the analytic expression of
$\tt0(\bfq)$. This reads
\begin{equation}\nonumber
\tt0(\bfq) = \prod_{j=1}^3\int_{-1}^1 e^{-iq_j \xi}d\xi = 8 \prod_{j=1}^3
\frac{\sin(q_j)}{q_j}.
\end{equation}
Since $\tt0(\pi\bfm)=0$ if $\bfm\ne {\bf 0}$, we can replace
${\tilde\Psi}^2(\pi\bfm,\alpha)$ with $(\tphi2(\pi\bfm,\alpha))^2$
in Eq.~(\ref{a6}) and finally write Eq.~(\ref{a2}) as
\begin{equation}\label{a6}
\biggl[\frac{\omega_{\sigma}}{2\pi}{\sum_{\bfm}}'
\frac{(\tphi2(\pi\bfm,\alpha))^2}
{\td_{\sigma}(\bfm)}\biggr] -
\int_{\C0}d^3\vxi\int_{\C0}d^3\vxi'
\frac{{\phi_0}^2(\vxi,\alpha){\phi_0}^2(\vxi',\alpha)}
{d_{\sigma}(\vxi - \vxi')}.
\end{equation}
Expression (\ref{2.30}) for $\upsilon_{\sigma}(\alpha)$ and
Eq.s~(\ref{2.31}) and (\ref{2.32}) for $\upsilon_{P,\sigma}(\alpha)$
and $\upsilon_{N,\sigma}(\alpha)$ immediately follow from Eq.~(\ref{a6})
recalling that Eq.~(\ref{a2}) is the content of the square brackets
in  Eq.~(\ref{2.25}).

\section{Asymptotic behaviour of $\epsilon_{\sigma}(r_s)$ at large $r_s$}
First of all Eq.s~(\ref{3.3}) and (\ref{3.5}) are derived as follows.
The condition that $\phi_0(\xi,\alpha)$ be normalized requires that
$G(\xi,\alpha)$ be normalized and this implies that
\begin{equation}\label{B.1}
C^{-2}(\alpha) = 2\int_0^{1}e^{-2\alpha x^2/(1-x^2)}dx.
\end{equation}
With the change of the integration variable: $x\to \sqrt{y}/\sqrt{1+y}$,
the above expression converts to
\begin{equation}\label{B.2}
C^{-2}(\alpha) = \int_0^{\infty} y^{-1/2}(1+y)^{-3/2}e^{-2\alpha y}dy.
\end{equation}
Recalling the general definition of the confluent hypergeometric function
$\Psi(a,b;z)$
\begin{equation}\label{B.3}
\Psi(a,b;z)\equiv\frac{1}{\Gamma(a)}\int_0^{\infty} e^{-xt}
t^{a-1}\,(1+t)^{c-a-1}\,dt,
\end{equation}
reported in \S 6.5 of Ref.\cite{ErdMaObTr}, from Eq.~(\ref{B.2}) immediately
follows that
\begin{equation}\label{B.4}
C^{-2}(\alpha) = \sqrt{\pi}\Psi(\frac{1}{2},\,0;\, 2\alpha),
\end{equation}
which is equivalent to Eq.~(\ref{3.3}). In the same way,
\begin{equation}\label{B.5}
2\int_0^1 \bigl[\partial_x\,G(x,\alpha)\bigr]^2dx =
2C^2(\alpha)\int_0^1 \bigl[\frac{2\alpha x e^{-\alpha x^2/(1-x^2)}}
{1-x^2}\bigr]^2dx
\end{equation}
and, by the previous change of the integration variable, one finds
\begin{eqnarray}
\quad \quad & &
2C^2(\alpha)\int_0^1 \bigl[2\alpha^2 e^{-2\alpha y}y^{1/2}
(1+y)^{3/2} \bigr]^2dy\nonumber\\
\quad & & = C^2(\alpha)\sqrt{\pi}\Psi(-\frac{3}{2},\, -2\, ,
\, 2\alpha)= \frac{\Psi(-\frac{3}{2},\, -2\, ,\, -2\alpha)}
{4\alpha\Psi(\frac{1}{2},\, 0\, ,\, 2\alpha)},\label{B.5a}
\end{eqnarray}
\ie\ Eq.~(\ref{3.5}).\ac
Finally we show that the SLQCS approximation implies
that $\epsilon_{\sigma}(r_s)$ at large $r_s$ behaves according to
Eq.~(\ref{3.12}). Once we have determined  $\alpha_{\sigma}(r_s)$, 
from  Eq.~(\ref{2.23}) follows that
\begin{equation}\label{B.6}
\epsilon_{\sigma}(r_s)= \frac{e^2}{2a_0}\biggl[
\frac{\tau_{\sigma}(\alpha(r_s))}{{r_s}^2} +
\frac{\upsilon_{{\sigma}}(\alpha(r_s))}{r_s}\biggr],
\end{equation}
while the $\alpha_{\sigma}(r_s)$s are determined solving the equations
\begin{equation}\label{B.7}
\frac{\partial}{\partial\alpha}\biggl(
\frac{\tau_{\sigma}(\alpha)}{r_s^2}
+\frac{\upsilon_{\sigma}(\alpha)}{r_s}\biggr)=0.
\end{equation}
Illustrating Fig.~2's results,  we already noted that
$\lim_{\alpha\to\infty}\upsilon_{\sigma}(\alpha)= - M_{dl,\sigma}$ and that
$\alpha \approx r_s^{1/2}$ at large $r_s$. Besides, based on the fact that
the  $\upsilon_{\sigma}(r_s)$s approach their limit values from the above,
 after putting $\beta\equiv 1/\alpha$ it appears reasonable to assume for
 the $\upsilon_{\sigma}(\alpha)$s, as $\beta\to 0^+$,  the following 
asymptotic behaviour
\begin{equation}\label{B.8}
 \upsilon_{\sigma}(\alpha) \approx -M_{dl,\sigma} + c_{\sigma}\beta +o(\beta),
\end{equation}
$c_{\sigma}$ being a positive constant. The asymptotic behaviour of
$\tau_{\sigma}(\alpha)$
at large $\alpha$ is easily obtained from that of $\Psi(a,,b;\,2\alpha)$,
reported in \S 6.13.1 of Ref.\cite{ErdMaObTr}. One finds that
\begin{equation}\label{B.9}
\tau_{\sigma}(\alpha)\approx \kappa_{\sigma}\biggl[\frac{1}{\beta}+\frac{3}{2}
+\frac{9\beta}{16}+o(\beta)\biggr].
\end{equation}
Eq.s~(\ref{B.8}) and (\ref{B.9}) allow us to evaluate the derivatives present
in Eq.~(\ref{B.7}) so that this equation, to the leading order, converts into
$-\kappa_{\sigma}/(\beta^2 r_s^2)+c_{\sigma}/r_s=0$. The solution is
\begin{equation}\label{B.10}
\alpha=\sqrt{c_{\sigma}r_s/\kappa_{\sigma}}.
\end{equation}
Its substitution in Eq.~(\ref{B.6}) yields the first two  leading terms of the
asymptotic expansion of $\epsilon_{\sigma}(r_s)$ at large $r_s$, ie\
\begin{equation}\label{}
\epsilon_{\sigma}(r_s)\approx \frac{e^2}{2a_0}\biggl[
-\frac{M_{dl,\sigma}}{r_s}+
\frac{\sqrt{\kappa_{\sigma}c_{\sigma}}}{r_s^{3/2}}+\cdots\biggr].
\end{equation}
This result shows that the SLQCS approach implies that, at very high dilution,
the leading term of the energy per particle is the classical Madelung value and
that the leading correction to this term decreases as $r_s^{-3/2}$ with a
positive
numerical factor equal to $\sqrt{\kappa_{\sigma}c_{\sigma}}$. Fig.2 results
indicate that this factor nearly equals  2.9 and 3.6 in the sc and bcc case,
respectively.

%%%%%\vfill\eject


\begin{thebibliography}{}
\bibitem{Fetter} A.L. Fetter and  J.D. Walecka,
  \emph{ Quantum Theory of Many-Particle Systems},
McGraw-Hill, New  York,  (1971).
\bibitem{Mahan} G.D. Mahan, \emph{Many-particle Physics},
    Plenum Press,  New York, (1981).
\bibitem{WigPR} E.P. Wigner, \emph{Phys. Rev.}, {\bf 46}, 1002, (1934).
\bibitem{Bloch} F. Bloch, \emph{Z. Physik} {\bf 57},  545, (1929).
\bibitem{Marder} M.P. Marder, \emph{Condensed Matter Physiscs}, Wiley, New York, (2000).
\bibitem{WigTrans} E.P. Wigner, \emph{ Trans. Farad. Soc.} {\bf 34}, 678, (1938).
\bibitem{Carr}  W.J. Carr, \emph{ Phys. Rev.} {\bf 122}, 1437, (1961).
\bibitem{Messiah} A. Messiah, \emph{M\'ecanique Quantique}, vol. II, Dunod, Paris, (1962).
\bibitem{FoulMiNeRa} W.M.C. Foulkes, L. Mitas, R.J. Needs and G. Rajagopal,\emph{ Rev. Mod. Phys.} {\bf 73}, 33,(2001).
\bibitem{Trail} J.R. Trail, M.D. Towler and R.J. Needs, \emph{Phys. Rev. B} {\bf 68}, 045107, (2003).
\bibitem{CeperAl} D.M. Ceperley and B.J. Alder, \emph{Phys. Rev. Lett.} {\bf 45},  567, (1980).
\bibitem{OrtiBa} G. Ortiz and P. Ballone, \emph{Phys. Rev. B} {\bf 50},  1391, (1994).
\bibitem{KwoCepMa} Y. Kwon, D.M. Ceperley and R.M. Martin, \emph{ Phys. Rev B} {\bf 58},  6800, (1998).
\bibitem{OrtHaBa} G. Ortiz, M. Harris and P. Ballone, \emph{ Phys. Rev. Lett.} {\bf 82},  5317, (1999).
\bibitem{Zong} F.H. Zong, C. Lin and D.M. Ceperley, \emph{Phys. Rev. E} {\bf 66}, 036703, (2002).
\bibitem{DruRaTrToNe} N.D. Drummond, Z. Radnai, J.R. Trail, M.D. Towler and R.J. Needs, \emph{Phys. Rev. B} {\bf 69}, 085116, (2004).
\bibitem{Kittel} C. Kittel, \emph{Introduction to Solid State Physics}, J. Wiley, New York, (2005).
\bibitem{Kohanoff} J. Kohanoff, \emph{Electronic Structure Calculations for
Solid and Molecules}, Cambridge Univ. Press, Cambridge, (2006), \S 3.1.
\bibitem{Slater} J.C. Slater, \emph{Insulators, Semiconductors and Metals.
Quantum Theory of Molecules and Solids}, (McGraw Hill, New York, 1967).
\bibitem{AshMer} N.W. Ashcroft, and N.D. Mermin, \emph{Solid State Physics},
Harcourt Coll. Pub., New York, 1976.
\bibitem{Cicca} S. Ciccariello, arXiv:0712.1463v1 [cond-matt.str-el] (2007).
\bibitem{ErdMaObTr} A. Erd\'eley, W. Magnus, F. Oberhettinger,  F.G. Tricomi,
     \emph{ Higher  Transcendental Functions I}, McGraw-Hill, New York, (1953).
%\bibitem{DysLen} F.J. Dyson and A. Lenard, \emph{J. Math. Phys.}{\bf 8}, 423, (1967).
%\bibitem{LenDys} A. Lenard and F.J. Dyson,\emph{J. Math. Phys.},{\bf 9},  698, (1968).
%\bibitem{LebLieb} J.L. Lebowitz and E.H. Lieb,\emph{ Phys. Rev. Lett.}, {\bf 22}, 631, (1969).
%\bibitem{Lieb}  E.H. Lieb,  \emph{ Rev. Mod. Phys}, {\bf 48}, 553, (1976).
%\bibitem{Erd} A. Erd\'eley, \emph{Asymptotic Expansions}, (Dover,  New York, 1956).
%\bibitem{CicDeC}S. Ciccariello and A. De Col, \emph{Eur. J. Phys.} {\bf 22},  629, (2001).
%\bibitem{Mack} W. Macke, \emph{Z. Naturforsch.}, {\bf 5a}, 192, (1950).
%\bibitem{GellMan} M. Gell-Mann and K.A. Brueckner, \emph{ Phys. Rev.},{\bf 106}, 364, (1957).
%\bibitem{OnsMiSt} L. Onsager, L. Mittag and M.J. Stephen,\emph{ Ann. Physik}, {\bf 18}, 71, (1966).
%\bibitem{Dubois} D.F. Du Bois, \emph{ Ann. Phys. (N.Y.)} {\bf 7}, 714, (1959).
%\bibitem{CarMar} W.J. Carr and A.A. Maradudin, \emph{ Phys. Rev.} {\bf 133}, A371, (1964).
%\bibitem{Endo} T. Endo, M. Moriuchi, Y. Takada and H. Yasuhara, \emph{Phys.Rev. B}, {\bf 59}, 7367, (1999).
%\bibitem{PerdZun} J.P. Perdew and A. Zunger, \emph{Phys. Rev. B} {\bf 23}, 5042, (1981).
%\bibitem{SenPa} G. Senatore and G. Pastore, \emph{Phys. Rev. Lett.} {\bf 64}, 303, (1990).
%\bibitem{MakiZo} K. Maki and X. Zotos, \emph{ Phys. Rev. B} {\bf 28}, 4349, (1983).
%\bibitem{Ferr} R. Ferrari, \emph{ Phys. Rev. B} {\bf 42}, 4598, (1990).
%\bibitem{BMStroc} F. Bagarello, G. Morchio and F. Strocchi, \emph{ Phys. Rev. B} {\bf 48},  5306, (1993).
\end{thebibliography}
\end{document}